\documentclass[smallextended]{svjour3}       
\smartqed  
\usepackage{graphicx}
\begin{document}

\title{The fallacy of Schott energy-momentum}
\titlerunning{The fallacy of Schott energy-momentum}
\author{Ashok K. Singal}
\authorrunning{A. K. Singal} 
\institute{A. K. Singal \at
Astronomy and Astrophysics Division, Physical Research Laboratory,
Navrangpura,\\ Ahmedabad - 380 009, India\\
\email{ashokkumar.singal@gmail.com}         }
\date{Received: date / Accepted: date}
\maketitle
\begin{abstract}
The incompatibility between Larmor's formula for radiation losses (at a rate proportional to square of  the acceleration of the electric charge) and the radiation reaction (the rate of loss of momentum of the accelerated charge proportional to its rate of change of acceleration) 
was recently shown to arise because a proper distinction is not kept between radiation losses calculated in terms of a retarded time and those expressed in terms of a ``real time''.
However, the occurrence of this disparity between two formulations is usually reconciled in literature by proposing an acceleration-dependent Schott energy lying somewhere in the nearby electromagnetic fields of an accelerated charge. But nobody has yet unambiguously demonstrated where the Schott energy actually lies in the fields. By scrutinizing electromagnetic fields of a uniformly accelerated charge, a mathematically tractable case, we show that contrary to the ideas prevalent in the literature, there is no evidence of any acceleration-dependent Schott energy-momentum in the electromagnetic fields, anywhere in the near vicinity of the charge or elsewhere. Accordingly, we expose the fallacy of the Schott energy-momentum term, which should henceforth be abandoned, in the electromagnetic radiation formulation.\\\\
\end{abstract}
\section{Introduction}
According to Larmor's formula (or its relativistic generalization, Li\'{e}nard's formula), electromagnetic power is radiated from an accelerated charge at a rate proportional to square of its acceleration \cite{la97,1,2,25}.
From that one can also infer the rate of momentum carried by the electromagnetic radiation which turns out to be directly proportional to the velocity vector of the charge multiplied by square of its acceleration \cite{ha95,51}.
The picture here does not seem to be complete however, because if one attempts to compute the consequential rate of energy-momentum loss from the radiating charge, one encounters not only a direct violation of the energy-momentum conservation law but also sees a conflict with the special theory of relativity \cite{68d}.

Abraham \cite{abr04,abr05} and Lorentz \cite{lor04,16} derived for an accelerated  charge, a formula for the self-force, widely known as radiation reaction, which gives the rate of loss of momentum of the accelerated charge proportional to its rate of change of acceleration \cite{7,24,3,20}. The same formula is also obtained,  independently, from momentum conservation law by using the Maxwell stress tensor to calculate the rate of electromagnetic momentum flow across a spherical surface surrounding the neighbourhood of a point charge \cite {68b}.
A scalar product of the radiation reaction with the velocity of the charge yields the rate of power loss of the accelerated charge. The radiative power loss can also be obtained directly from the Poynting flux in the neighbourhood of a point charge in arbitrary motion, leading exactly to the same formula \cite {68a}.

The disparity between the two power loss formulas (one proportional to the square of acceleration and the other proportional to the scalar product of the velocity and the rate of change of acceleration of the charge), has remained a nagging puzzle for almost a century. It was recently shown that this well-known incompatibility in the two formulas is succinctly resolved when a proper distinction is made between radiation losses expressed in terms of a retarded time and those expressed in terms of a ``real time'' \cite {68}. 
However, according to the conventional wisdom, the radiative power loss is given correctly by Larmor's formula, while the rate of loss of momentum is described correctly by the radiation reaction formula, and these two apparently conflicting results are reconciled by proposing the presence of an extra term, called Schott term \cite{7}, in the fields of an accelerated charge. However, a physical meaning of this acceleration-dependent Schott term is still not clear \cite{51,6,44,eg06,41,56,row10} and one does not encounter such an acceleration-dependent energy-momentum term elsewhere in physics. 

We shall first get the Schott term in a 4-vector form, from the differences in the two conflicting formulas. The Schott energy-momentum is thought to 
be present in the electromagnetic fields in the vicinity of an accelerated charge. We shall here examine the case of a uniformly accelerated charge, where the expression for the electromagnetic fields is relatively simple and the Schott term, if present therein, should be tractable mathematically in an exact manner. From a careful scrutiny of the electromagnetic fields of such an accelerated charge, we examine  whether there is an evidence of the Schott energy-momentum term in the vicinity of the charge as postulated in the literature. 
\section{Larmor's/Li\'{e}nard's radiation formula}
In Larmor's/Li\'{e}nard's formulation, the rate of energy-momentum loss of an accelerated charge due to {\em radiation damping}, expressed in a 4-vector form, is
\begin{equation}
\label{eq:21b1}
{\cal F}^{\mu}_1=\frac{2e^{2}}{3c^{5}}\:\dot{v}^{\alpha}\dot{v}_{\alpha}{v}^{\mu} \:,
\end{equation}
where in all covariant equations, dot represents a proper time derivative \cite{51}.

The time part, ${\cal F}^{0}_1$, is $\gamma {\cal P}_{1}/c$, where
\begin{equation}
\label{eq:11b}
{\cal P}_{1}=\frac{2e^{2}}{3c^{3}} \:\dot{v}^{\alpha}\dot{v}_{\alpha}
=\frac{2e^{2}\gamma ^{4}}{3c^{3}}\left[\dot{\bf v}\cdot\dot{\bf v} +
\frac{\gamma^2(\dot{\bf v} \cdot{\bf v})^2}{c^2}\right]
\end{equation} 
is Li\'{e}nard's formula (in cgs units) for power being lost by a radiating charge \cite{1,2,25}, 
and the space part, ${\cal F}^{i}_1$, is $\gamma$ times the $i$th component, for $i=1,2,3$, of the rate of momentum being lost into radiation \cite{ha95,51}
\begin{equation}
\label{eq:21b}
{\bf F}_{1}=\frac{{\cal P}_{1}}{c^2}{\bf v}\:.
\end{equation}

The radiative power loss for a charge moving with a  non-relativistic velocity ($v\ll c$), accordingly, is \cite{1,2,25}
\begin{equation}
\label{eq:21a}
{{\cal P}_{1}}=\frac{2e^2\dot{\bf v}^{2}}{3c^3} \:,
\end{equation}
while the net rate of momentum loss to radiation by such a charge is nil 
\begin{equation}
\label{eq:21a1}
{\bf F}_{1}=0\:.
\end{equation}
This is consistent with the radiation pattern possessing an azimuth symmetry  ($\propto \sin^2\phi$) in the case of a non-relativistic motion \cite{1,2,25}.
\section{Abraham-Lorentz radiation reaction}
Larmor's formula purportedly uses Poynting's theorem of energy conservation to relate Poynting flux through a spherical surface of radius $r$ at a time $t$, to the rate of loss of kinetic energy of the radiating charge at a retarded time $t-r/c$. 
However, in Poynting's theorem {\em all quantities} are supposed to be calculated for the {\em same instant of time} \cite{1,2,25}. A correct application of the Poynting's theorem, using real-time values of the charge motion, gives 
instantaneous power loss of the charge (in a non-relativistic motion) as \cite{68a}
\begin{equation}
\label{eq:6.1}
{\cal P}_2 =-\frac{2e^{2}}{3c^{3}}\ddot{\bf v}\cdot{\bf v}\:.
\end{equation}

Similarly the electromagnetic momentum flow across a surface surrounding the vicinity of a point charge, computed employing the Maxwell's stress tensor, yields a rate of {\em loss of mechanical momentum} of the charge \cite{68b}
\begin{equation}
\label{eq:3a}
{\bf F}_2= -\frac{2e^{2}}{3c^{3}}\ddot{\bf v}\:.
\end{equation}
Equation~(\ref{eq:3a}) is the famous Abraham-Lorentz radiation reaction formula, derived usually in a quite involved way by computing the net self-force on the accelerated charge \cite{abr04,abr05,lor04,16,7,24,3,20}. For this one considers the charge to be in the shape of a small spherical shell and the force 
on every tiny bit of the shell, due to the time-retarded fields of the rest of the shell charge distribution, is calculated and then the total force is calculated by summing over the whole spherical shell.
But the same result now has been obtained in an independent manner from the  momentum conservation theorem \cite{68b}.

A relativistic generalization of Eq.~(\ref{eq:6.1}) yields  \cite{20,68b}
\begin{equation}
\label{eq:10}
{\cal P}_2= -\frac{2e^{2}\gamma ^{4}}{3c^{3}}\left[\ddot{\bf v}\cdot{\bf v}+
\frac{3\gamma ^{2}(\dot{\bf v}\cdot{\bf v})^{2}}{c^{2}}\right]\:,
\end{equation}
while a relativistic generalization of Eq.~(\ref{eq:3a}) is \cite{3,20,68b}
\begin{eqnarray}
\nonumber
{\bf F}_2= -\frac{2e^{2}\gamma^2}{3 c^{3}}\left[ \ddot{\bf v}+\frac{\gamma ^{2}(\ddot{\bf v}\cdot{\bf v}){\bf v}}{c^{2}}
+\frac{3\gamma ^{2}(\dot{\bf v}\cdot{\bf v})\dot{\bf v}}{c^{2}}\right.\\
\label{eq:3i}
\left.+\frac{3\gamma ^{4}(\dot{\bf v}\cdot{\bf v})^2{\bf v}}{c^{4}}\right].\:
\end{eqnarray}

We can express Eqs.~(\ref{eq:10}) and (\ref{eq:3i}) in a 4-vector form 
${\cal F}^{\mu}_2$, where  
\begin{eqnarray}
\label{eq:10.1}
{\cal F}^{0}_2&=&\frac{\gamma {\cal P}_2}{c},\\
{\cal F}^{i}_2&=&\gamma {\bf F}^i_2,\;\;\;\; i=1,2,3\:.
\end{eqnarray}
\section{Schott energy-momentum term}
Power loss by the charge due to radiation reaction (Eq.~(\ref{eq:6.1})), is related to the radiated power by Larmor's formula (Eq.~(\ref{eq:21a})), in a  non-relativistic case, as 
\begin{equation}
\label{eq:9.1}
{\cal P}_2={\cal P}_1-\frac{2e^{2}}{3c^{3}}\frac{{\rm d}(\dot{\bf v}\cdot{\bf v})}{{\rm d}t}\:.
\end{equation}
The last term on the right hand side in Eq.~(\ref{eq:9.1}) is known as the Schott term, after Schott \cite{7} who first brought it to attention. Schott term is a total derivative  and is thought in literature to arise from an acceleration-dependent energy, $-{2e^{2}}(\dot{\bf v}\cdot{\bf v})/{3c^{3}}$, in electromagnetic fields
\cite{6,44,eg06,41,56,row10}.

We can also express ${\bf F}_2$ in terms of ${\bf F}_1$ (Eqs.~(\ref{eq:3a}) and (\ref{eq:21a1}), again  in a non-relativistic case) as
\begin{equation}
\label{eq:9.1a}
{\bf F}_2={\bf F}_1-\frac{2e^{2}}{3c^{3}}\ddot{\bf v}\:.
\end{equation}
The last term on the right hand side, again, is a total derivative, assumedly arising from an acceleration-dependent momentum, 
$-{2e^{2}}\dot{\bf v}/{3c^{3}}$, apparently in electromagnetic fields.

Radiation reaction in the covariant form \cite{lau09} yields a 4-vector, ${\cal F}^\mu_s$, for the Schott term 
\begin{equation}
\label{eq:9.1b}
{\cal F}^\mu_2={\cal F}^\mu_1+{\cal F}^\mu_s={\cal F}^\mu_1-\frac{2e^{2}}{3c^{3}}\ddot{v}^\mu\:.
\end{equation}
${\cal F}^\mu_s$ is a proper-time derivative of the Schott energy-momentum, ${\cal E}^\mu_s=-{2e^{2}}\dot{ v}^\mu/{3c^{3}}$. The 4-acceleration $\dot{ v}^\mu$ is obtained from the 4-velocity ($\gamma c,\gamma \bf v$) by a differentiation with proper time and the 4-vector ${\cal E}^\mu_s$ then is
\begin{eqnarray}
\label{eq:9.1c}
{\cal E}^0_s&=&-\frac{2e^{2}\gamma^4}{3c^{4}}{\dot{\bf v}\cdot{\bf v}},\\
\label{eq:9.1c1}
{\cal E}^i_s&=&-\frac{2e^{2}\gamma^2}{3c^{3}}\Big[\dot{\bf v}^i+\frac{\gamma^2({\bf v}\cdot\dot{\bf v} ){\bf v}^i}{c^2}\Big],\;\;\;\; i=1,2,3\:.
\end{eqnarray}
Equation~(\ref{eq:9.1b}) can now be explicitly verified by a proper-time differentiation of ${\cal E}^\mu_s$, in conjunction with Eqs.~(\ref{eq:11b}), (\ref{eq:21b}), (\ref{eq:10}) and (\ref{eq:3i}), to give 
\begin{eqnarray}
\label{eq:10.1}
{\cal F}^0_s=-\frac{2e^{2}}{3c^{3}}\ddot{v}^0&=&\frac{\gamma ({\cal P}_2-{\cal P}_1)}{c},\\
{\cal F}^i_s=-\frac{2e^{2}}{3c^{3}}\ddot{v}^i&=&\gamma ({\bf F}^i_2-{\bf F}^i_1)\;\;\;\; i=1,2,3\:.
\end{eqnarray}

It may be noted that in case of radiation reaction, power and force  (Eqs.~(\ref{eq:10}) and (\ref{eq:3i})) are related by ${\cal P}_2={\bf F}_2.{\bf v}$, implying 
\begin{eqnarray}
\label{eq:10.11}
{\cal F}^0_2=\frac{{\cal F}^i_2 v_i}{\gamma c}\:,
\end{eqnarray}
where $v_i$ ($i=1,2,3$) stands for $i$th component of the 4-velocity. 
In contrast, the relation between force and power in case of Larmor's radiation formula, ${\bf F}_{1}={{\cal P}_{1}}{\bf v}/{c^2}$ (Eq.~(\ref{eq:21b})), implies
\begin{eqnarray}
\label{eq:10.12}
{\cal F}^0_1\ne\frac{{\cal F}^i_1 v_i}{\gamma c}\:.
\end{eqnarray}
Also 
\begin{eqnarray}
\label{eq:10.14}
{\cal F}^0_s\ne\frac{{\cal F}^i_s v_i}{\gamma c}\:,
\end{eqnarray}
because
\begin{eqnarray}
\label{eq:10.15}
\ddot{v}^0=\frac{(\ddot{v}^i v_i+\dot{v}^\mu \dot{v}_\mu)}{\gamma c}\:.
\end{eqnarray}

An uncomfortable question in Larmor's radiation loss formula arises in case of an accelerated charge in its instantaneous rest frame. 
Due to zero velocity of the charge, it 
could not lose any kinetic energy into radiation. However, Larmor's formula, according to which the radiated
power is proportional to square of acceleration of the charge, 
yields a finite power loss. 
Even if the external force causing the acceleration of the charge, were considered to be responsible for the radiative power as well, it could not have done so, 
because at that instant the rate of work being done by the external force would also be zero as the system has a zero velocity.
This embarrassing question is purportedly resolved by proposing that there is an equivalent decrease in the Schott energy (an acceleration-dependent internal energy!) stored within the electromagnetic fields in the close vicinity of the accelerated charge. According to this argument, even if the Schott energy term may be zero in the instantaneous rest-frame (Eq.~(\ref{eq:9.1c})), its temporal derivative (Eq.~(\ref{eq:10.1})) yields a finite power loss for the instantly stationary charge equal to that expected from Larmor's formula (Eq.~(\ref{eq:21a})).

However, even if this might seem to resolve the particular energy conservation problem, it gives rise to another awkward question about the presence of momentum for an instantly stationary charge. From Eq.~(\ref{eq:9.1c1}) we infer that there is a finite momentum, $-{2e^{2}}\dot{\bf v}/{3c^{3}}$, in electromagnetic fields in the vicinity of the charge, even at the instant when the charge is  stationary $({\bf v}=0)$. Now this apparent momentum, which is directly proportional to the acceleration of the charge, and is strangely independent of the velocity of the charge, at least in the non-relativistic case, raises a vexing question -- How come there is supposedly a finite momentum in the fields of a stationary charge when there is {\em no motion of any kind} at that instant? We want to examine the case of a uniformly accelerated charge, where it may be possible to tract the question in exact mathematical details whether the  electromagnetic fields really harbour the Schott energy-momentum, somewhere in the close vicinity of such a charge, as opined in the literature \cite{eg06,41}.
\section{Electromagnetic fields around the ``present'' position of a uniformly accelerated charge -- no trace of Schott energy-momentum anywhere}
A uniformly accelerated motion is understood to imply a motion with a constant proper acceleration, say, ${\bf g}$. 
For simplicity, we may assume a one-dimensional motion, ${\bf v \parallel g}$,  since, using a Lorentz transformation, we can always switch to an appropriate inertial frame in which the velocity component perpendicular to the acceleration vector is zero. 
In fact, at any given event, transforming to the instantaneous rest frame  guarantees that for a constant proper acceleration the motion will be along one dimension.

In the case of a uniform proper acceleration ${\bf g}$, Schott term, ${\cal F}^\mu_s$  is the 4-vector
\begin{eqnarray}
\label{eq:9.1c2}
{\cal F}^0_s&=&-\frac{2e^{2}}{3c^{4}}\gamma{\bf g}^2,\\
\label{eq:9.1c3}
{\cal F}^i_s&=&-\frac{2e^{2}}{3c^{5}}\gamma{\bf g}^2{\bf v}^i,\;\;\;\; i=1,2,3\:,
\end{eqnarray}
while from Eqs.~(\ref{eq:9.1c}) and (\ref{eq:9.1c1}), the Schott energy-momentum 4-vector ${\cal E}^\mu_s$ is given by
\begin{eqnarray}
\label{eq:9.1c4}
{\cal E}^0_s&=&-\frac{2e^{2}}{3c^{4}}\gamma{\bf g}\cdot{\bf v},\\
\label{eq:9.1c5}
{\cal E}^i_s&=&-\frac{2e^{2}}{3c^{3}}\gamma\:{\bf g}^i,\;\;\;\; i=1,2,3\:.
\end{eqnarray}
As it was mentioned earlier, this Schott energy-momentum is thought to be present in the electromagnetic fields in the near vicinity of an accelerated charge. 

In order to explore the electromagnetic fields around the charge in its neighbourhood, we could attempt to express fields at a time $t$ with respect to the  position and motion of the charge also at the same instant $t$. It may not be quite  feasible to do so for an arbitrary motion of the charge. However, for a uniformly  accelerated charge, it is possible to solve the expression for electromagnetic fields not necessarily in terms of motion of the charge at retarded time, instead wholly in terms of the ``real-time'' motion of the charge \cite{32}. 

We assume that the constant proper  acceleration vector, ${\bf g} \;(=\gamma^{3} \dot{\bf v})$, is along the  
$+z$ axis and that the charge, coming 
from $z=\infty$ at time $t=-\infty$, initially moves along $-z$ direction, getting constantly decelerated till it
comes to rest momentarily at a point $z=\alpha$ at time $t=0$, and 
then onwards moves with an increasing speed along the $+z$ direction.
Without any loss of generality, we can choose the origin of the coordinate system so that $\alpha=c^{2}/g$, then the position and velocity of the charge at a time $t$ are given by $z_{c}=(\alpha^{2}+c^{2}t^{2})^{1/2}$ and $\beta=v/c=ct/z_{c}$. 
The charge happens to be at the same point on the $z$-axis at times $-t$ and $t$, but with velocity in opposite directions, i.e., ${\mbox{\boldmath $\beta$}}(t)=-{\mbox{\boldmath $\beta$}}(-t)$ 

The expression for electromagnetic fields of such a charge \cite{5}, 
in a spherical coordinate system ($r,\theta,\phi$) with origin 
at the instantaneous charge position \cite{18}, is
\begin{eqnarray}
E_{r} &=&\frac{e(1+\eta \cos\theta)}{r^{2}\gamma^{2}
(1+2\eta\cos\theta+\eta^{2}-\beta^{2}\sin^{2}\theta)^{3/2}}
\nonumber\\
E_{\theta}&=&\frac{e\eta \sin\theta}{r^{2}\gamma^{2}
(1+2\eta\cos\theta+\eta^{2}-\beta^{2}\sin^{2}\theta)^{3/2}}
\nonumber\\
\label{eq:32a1}
B_{\phi}&=&\frac{e\beta \sin\theta}{r^{2}\gamma^{2}
(1+2\eta\cos\theta+\eta^{2}-\beta^{2}\sin^{2}\theta)^{3/2}}
\end{eqnarray}
with $\eta=gr/2\gamma c^{2}$. All the remaining field components are zero.

Since the magnetic field $B_{\phi}$ has linear dependence on $\beta$ (Eq.~(\ref{eq:32a1})), therefore at any given location ($r,\theta,\phi$), $B_{\phi}(t)=-B_{\phi}(-t)$. On the other hand, the electric field components $E_{r},E_{\theta}$ do not have such linear dependence on $\beta$ and ${\bf E}(t)={\bf E}(-t)$. 

From Eq.~(\ref{eq:32a1}) we can infer the following:\\
(i) For $g=0$, $\eta=0$ and in that case the fields reduce to that of a charge moving with a uniform velocity ${\mbox{\boldmath $\beta$}}$, with the electric field everywhere in a radial direction from the present position of the charge with ${\bf B}= {\mbox{\boldmath $\beta$}}\times {\bf E}$ \cite{1,2,25}.\\
(ii)  For a finite $g$ ($\eta \ne 0$), the radial component of the Poynting vector, ${\cal S}_r = c(E_\theta B_\phi)/{4\pi}$, at any time $t<0$, when the charge is getting decelerated, is everywhere (i.e., at any field point $r,\theta,\phi$) {\em pointing inward}, toward the present position of the charge.\\
(iii) At $t=0$, $\beta=0$, implying ${\bf B}=0$ everywhere. Thus there is no Poynting vector seen anywhere at $t=0$. \\
(iv) At time  $t>0$, when the charge is accelerating, the radial component of the Poynting vector is everywhere pointing away from the present position of the charge. In fact, everywhere the Poynting vector at time $t_1$ is equal and opposite to that at time $-t_1$, for all $t_1$ values. 

The electromagnetic field energy in a volume $\cal V$ is given by the volume integral
\begin{equation}
\label{eq:32a3}
\frac{1}{8\pi}\int_{\cal V} {{\rm d}V} \;({E^{2}+B^{2}})\:.
\end{equation}
The field energy density, $({E^{2}+B^{2}})/{8\pi}$, being equal at times $t_1$ and $-t_1$, its volume integral over {\em any chosen} $\cal V$, whether in the vicinity of the charge or in some far-off zone, is also equal at times $t_1$ and $-t_1$. 
Now, the acceleration-dependent Schott energy term, according to Eq.~(\ref{eq:9.1c4}), is equal but opposite at $t_1$ and $-t_1$ (because $v=-c^{2}t_1/z_{c}$ at $-t_1$). Thus the Schott energy should be making a positive contribution at $-t_1$ and a negative contribution at $t_1$, which is not consistent with the fact that the actual field energy, computed from Eq.~(\ref{eq:32a3}), is identical at $t_1$ and $-t_1$.

One can also compute the electromagnetic field momentum contained within a volume $\cal V$  from
\begin{equation}
\label{eq:1f1b}
\frac{1}{4\pi c}\int_{\cal V}{{\rm d}V}\:({\bf E}\times{\bf B}).
\end{equation}
Since $\bf B=0$ at $t=0$ (Eq.~(\ref{eq:32a1})), there is no momentum in the electromagnetic fields anywhere, whether in the vicinity of the charge or in the far-off regions, in the instantaneous rest frame. Therefore Eq.~(\ref{eq:9.1c5}) is clearly violated where the Schott momentum is proportional to $-\bf g$ even at $t=0$, when the charge is instantly stationary. 

Further, for  $t\ne0$, from Eq.~(\ref{eq:1f1b}) in conjunction with Eq.~(\ref{eq:32a1}), the electromagnetic field momentum is not only equal but in opposite directions at times $t_1$ and $-t_1$, it is also directly proportional to the instantaneous velocity $\beta$. 
Now, this again is not in agreement with Eq.~(\ref{eq:9.1c5}), where not just the magnitude but also the direction of the Schott momentum vector should remain the same at times $t_1$ and $-t_1$, opposite in direction but directly proportional to the acceleration $g$, unlike the electromagnetic field momentum that is directly proportional to the instantaneous velocity $\beta$. In fact, a finite Schott momentum for an accelerated charge in its instantaneous rest frame, as inferred from Eq.~(\ref{eq:9.1c5}), would from the strong principle of equivalence \cite{4} imply a finite momentum ${2e^{2}}g/{3c^{3}}$ associated with a charge that is continually at rest in a gravitational field of strength $g$, an unpalatable inference for an otherwise completely static system.

Thus we find no signature of the acceleration-dependent Schott energy-momentum terms that were in accordance with Eqs.~(\ref{eq:9.1c4}) and (\ref{eq:9.1c5}). 
We may add here that the introduction of the Schott energy term to account for the power loss into radiation but without any equivalent rate of decrease of kinetic energy of the radiating charge, say, in the instantaneous rest frame, is akin to  the proposal of the loss of internal (rest mass!) energy \cite{8} without a loss of momentum (c.f. Eqs.~(\ref{eq:21a}) and (\ref{eq:21a1})), though in the case of Scott energy it is thought to be an acceleration-dependent extraneous kind of energy (neither the rest mass energy nor the kinetic energy, not even some kind of potential energy that may depend upon location in an external field) present in the electromagnetic fields and which does not seem to make an appearance elsewhere in physics. In any case, we see no evidence of the presence of such an energy term in the fields of a uniformly accelerated charge. 
Actually it has recently been shown that the Schott term is merely a difference in rate of change of energy in self-fields of the charge between retarded and real times \cite{68,68c,68f}.
and contrary to the ideas that have been proposed in the literature  \cite{51,6,44,eg06,41,56,row10}, there is no acceleration-dependent extra energy term lurking somewhere in the electromagnetic fields whether in the near vicinity of the charge or elsewhere.
\section{Conclusions}
From the difference between Larmor's/Li\'{e}nard's radiation formula and Abraham-Lorentz radiation reaction formula, we arrived at the expression for Schott energy-momentum for an accelerated charge. We demonstrated that in the electromagnetic fields of a uniformly accelerated charge there is no evidence, whatsoever, of the Schott energy-momentum terms, whether in the near vicinity of the charge or elsewhere. The presence of such terms would have been, even otherwise, in conflict with the strong principle of equivalence as one would then infer from them, among other things, a finite momentum for a charge continually at rest in a gravitational field, an unpalatable inference for an otherwise completely static system. Since   the difference between two formulations is resolved when a proper distinction is made between radiation losses calculated in terms of a retarded time and those expressed in terms of a ``real time'', the proposition of Schott energy-momentum terms in fields is superfluous and needs to be abandoned.
{}
\end{document}